**The Effect of Residual Confounding in the Analysis of Large Observational Data Sets**

**S. Stanley Young[1] and Warren B. Kindzierski[2]**

[1] CGStat, Raleigh, NC, USA
[2] Independent consultant, St Albert, Alberta, Canada
Correspondence: S. Stanley Young, Email: genetree@bellsouth.net

**Summary**
We review how residual confounding in large observational data sets can mistakenly lead researchers to declare a predictor variable statistically significant. As sample sizes become very large, standard errors decrease dramatically, causing unmeasured variables correlated with the predictor of interest to overwhelm the model and yield precise estimates of the wrong thing, the residual confounder. Three examples from environmental epidemiology are provided. Specification Curve Analysis (SCA) and Local Control Analysis (LCA) are offered to reveal model selection bias and true null effects.

**Introduction**
In 2011, Young & Karr examined randomized trials that tested research claims coming from twelve large observational epidemiology studies. A total of 52 claims were examined, none of which were replicated. This study was part of the claimed replication crisis (Baker, 2016). It had been noted by Young & Karr (2011) that the observational studies examined many questions, and that statistical adjustments were not made for multiple testing.

Multiple researchers—Young & Karr (2011), Begley & Ellis (2012), Gelman & Loken (2014)—have noted that there are many modeling choices available in observational research and that these choices are seldom considered in the evaluation of analysis results. Finally, a survey published in the journal *Nature* of 1,576 researchers found that 52% of those surveyed agree that there is a significant ‘crisis’ of reproducibility (Baker 2016).

Two relatively new analysis methods are useful for understanding the problems and limitations related to the analysis of large observational data sets – Specification Curve Analysis, SCA (Simonsohn et al. 2020), and Local Control Analysis, LCA (Obenchain and Young 2013, 2017). SCA is used to evaluate modeling choices. Many reasonable analyses are proposed, and all are computed in SCA. Here, researchers thought that these results would converge to the “correct” answer, but in many SCAs that did not happen. Across different SCA studies, results ranged widely from statistical significance negative to statistical significance positive and all results in between.

Local Control Analysis, LCA, clusters the data set on important covariates, not using the outcome or predictor of interest, and computes a simple analysis within each cluster, either A vs B, simple linear regression or ranked correlation. The analysis results are then examined to determine if results vary across clusters. Also, of interest is whether the results are consistent across clusters or whether they vary depending on the variables used in clustering? Our experience is that the relationship, for example, between a predictor, X, and an outcome, Y, vary across clusters, but can be largely explained by the covariates (Young et al. 2015, Obenchain and Young 2015, and Obenchain et al. 2019). Refer to Endnote 1.

## The Model

Residual confounding can happen in multiple linear regression when a term in the "Error" component is correlated with the predictor and mistakenly leads to the predictor being declared "significant" (see the model below). The potential problem becomes more pronounced with large data sets. As sample size (N) increases, the Standard Error (SE) decreases so that the residual confounding variable or variables overwhelms the predictor, when the effect of the predictor is small or null.

The model for multiple linear regression is:

$Y = b_0 + b_1 P_1 + \Sigma b_i C_i + \Sigma b_i U_i + \textit{Error}$, where

  Y is a prediction,

  $P_1$ is the predictor of interest,

  $C_i$ are covariates to be statistically removed from the prediction, and

  *Error* is what is left over.

*Error* is often assumed to be normally distributed with mean zero.

  $= \Sigma b_i U_i$, where the $U_i$ are (unknown or unmeasured) variables that are not in the formal model.

## Big N

Recently, sample sizes of observational data sets have increased dramatically. It is not uncommon to have data sets of tens of thousands to millions. As N increases, the Standard Error decreases. The Mean Square Estimator of a parameter is equal to the Bias of the estimator squared plus the Variance of the estimator.

$$\mathrm{MSE}(\hat{\theta}) = \mathrm{Bias}(\hat{\theta})^2 + \mathrm{Var}(\hat{\theta})$$

But if there is bias, it does not decrease as sample size increases.

It is possible for variables correlated with a predictor of interest that are not in the model to flow through the analysis and be mistakenly taken as the effect of the variable of interest. This mistake is compounded by a small SE. It can end up that the predictor is null, and the residual confounder (correlated with the predictor) measured with great precision is taken as the effect of the predictor.

In other words… one can end up with a very precise estimate of the wrong thing, the residual confounder!

## Examples/Results

Three example meta-analysis (MA) studies were found that deal with associations between various air components and health effects:

- one for mortality (Orellano et al. 2020)
- two for asthma exacerbations (Zheng et al. 2015 and Orellano et al. 2017)

Each MA example presents the combined statistic from the MA's multiple individual base studies (i.e., a pooled relative risk) and 95% lower and upper confidence limits. In the case of Zheng et al. (2015), a measure of base study heterogeneity, $I^2$, was given.

**Orellano et al. (2020) meta-analysis of air quality–all-cause death associations (196 studies):**

| Air component | Pooled Relative Risk (RR) | Lower 95% Confidence Limit | Upper 95% Confidence Limit |
|---|---|---|---|
| PM10 | 1.0041 | 1.0034 | 1.0049 |
| PM2.5 | 1.0065 | 1.0044 | 1.0086 |
| NO2 | 1.0072 | 1.0059 | 1.0085 |
| O3 | 1.0043 | 1.0034 | 1.0052 |

**Zheng et al. (2015) meta-analysis of air quality–asthma exacerbations (87 studies):**

| Air component | Pooled Relative Risk (RR) | Lower 95% Confidence Limit | Upper 95% Confidence Limit | $I^2$ (%) |
|---|---|---|---|---|
| O3 | 1.009 | 1.006 | 1.011 | 87.8 |
| CO | 1.045 | 1.029 | 1.016 | 85.7 |
| NO2 | 1.018 | 1.014 | 1.022 | 87.6 |
| SO2 | 1.011 | 1.007 | 1.015 | 77.1 |
| PM10 | 1.010 | 1.008 | 1.013 | 69.1 |
| PM2.5 | 1.023 | 1.015 | 1.031 | 82.8 |

**Orellano et al. (2017) meta-analysis of air quality–asthma exacerbations (22 studies –** database searches retrieved 208 records, and 22 studies were selected for quantitative analysis**):**

| Air component | Pooled Relative Risk (RR) | Lower 95% Confidence Limit | Upper 95% Confidence Limit |
|---|---|---|---|
| NO2 | 1.024 | 1.005 | 1.043 |
| SO2 | 1.039 | 0.988 | 1.094 |
| PM10 | 1.024 | 0.998 | 1.053 |
| PM2.5 | 1.028 | 1.009 | 1.047 |
| CO | 1.045 | 1.005 | 1.086 |
| O3 | 1.032 | 1.005 | 1.060 |

**Discussion**

For the Orellano et al. (2020) MA of air quality–all-cause death associations, the pooled RRs are very small and near no effect, i.e., a Relative Risk of 1.0000. The Confidence Limits are very narrow, indicating a small SE. Also note that pooled RRs are small with very tight confidence limits for all three MA studies. All pooled statistics are quite similar for air components within a study, which seems implausible given different predictors (air components).

For the Zheng et al. (2015) MA, the within study base study heterogeneity is large, indicating across study heterogeneity. For example, Higgens et al. (2026) note that a rough guide for interpretation of heterogeneity (inconsistency) in statistics pooled in the context of meta-analyses of randomized trials is: considerable heterogeneity for $I^2$ ranging from 75 to 100%. Note that by looking at all the components together, strength is borrowed to help interpret/evaluate separate components.

There are many small factors correlated with death (Wang et al. 2024). There are many small factors correlated with asthma (Kindzierski et al. 2021). In both cases, these small factors are not used in modeling; they are residual factors. These results are consistent with Residual Bias being the driving force in the analyses, not the air component.

**Reasonable Thesis**

All the air components on the important list in these analyses are plausible true nulls [*or so insignificant (trivial) that they are effectively nulls in the presence of Residual Bias, large N, and small SE*]. It makes sense that if the results are due to residual confounding that the RRs of all air components within a MA study would be similar (i.e., it is small residual confounding that is giving rise to the observed RRs).

**Recommendation for Residual Confounding**

The residual confounding argument depends on the three meta-analyses, which in turn depend on their base studies. It would be useful to select, at random, a 10 to 15 percent sample of the base studies and evaluate their reliability using SCA and LCA. We expect SCA to return very variable results consistent with the air component effects being null. In addition, we expect LCA to return null or very near null results that are governed by the covariates used in clustering – a result obtained for reanalysis of a Pope et al. (2009) data set on the association between PM2.5 and longevity by Young et al. (2015).

**Endnotes**

1. The measured effect, Y|X, inside each cluster, hypercube, where the cluster variables are within tight ranges. So, the cluster variables have little leverage. The statistic coming out of the cluster is a within block value. Obenchain assigns all the objects in a cluster, the cluster value, say the non-parametric rank correlation. Next, Recursive Partitioning (RP) is applied to all the values coming from the blocks. RP shows how the Local Rank Correlation (LRC) changes as function of the covariates (those used in clustering and the rest). There are interactions amongst the LRCs. Our experience is that the RP explains a high percentage of the differences among LRCs. The covariates "explain" the within cluster effects. RP is improved in the sense that the Y is free of covariate effects and sample size is not used to deal with getting rid of covariate effects. LCA separates "pure" Y|X effects from covariate effects. It reveals the treatment by cluster (block) interaction. Local Control Analysis is relatively new. The reader is directed to Obenchain (2013, 2017) for further information.